\documentstyle[aps,pre,psfig]{revtex}

\begin{document}
\draft

\twocolumn[\hsize\textwidth\columnwidth\hsize\csname @twocolumnfalse\endcsname

\title{Water Droplets in a Spherically Confined Nematic Solvent:\\
  A Numerical Investigation}

\author{Holger Stark and Joachim Stelzer}
\address{Institut f\"ur Theoretische und Angewandte Physik, Universit\"at
Stuttgart, Pfaffenwaldring 57, D-70550 Stuttgart, Germany}
\author{Ralf Bernhard}
\address{IMPACT Me{\ss}technik GmbH, Maybachstr.~25,
D-71332 Waiblingen, Germany}

\maketitle

\begin{abstract}

Recently, it was observed that water droplets suspended
in a nematic liquid crystal form linear chains 
(Poulin {\em et al.}, Science {\bf 275},  1770  (1997)). 
The chaining occurs, {\em e.g.}, in a large nematic drop with
 homeotropic boundary conditions at all the surfaces.
Between each pair of water droplets a point defect in
the liquid crystalline order was found in accordance with
topological constraints. This point defect causes a repulsion between the
water droplets.

In our numerical investigation we limit ourselves to a chain of 
two droplets. For such a complex geometry we use the method of finite 
elements to minimize the Frank free energy. We
confirm an experimental observation that the distance $d$ of the
point defect from the surface of a water droplet
scales with the radius $r$ of the droplet like  $d \approx 0.3 \, r$.
When the water droplets are moved apart, we find that
the point defect does not stay in the middle between the droplets,
but rather forms a dipole with one of them. This confirms a
theoretical model for the chaining. Analogies to a second order phase
transition are drawn. We also find the dipole when one
water droplet is suspended in a bipolar nematic drop with two boojums,
{\em i.e.}, surface defects at the outer boundary. 
Finally, we present a configuration where
two droplets repel each other without a defect between them.
\end{abstract}


\vskip2pc]

\narrowtext

\section{Introduction} \label{sec intro}
Liquid crystal research is, without any doubt, strongly driven by its
importance for technological applications, namely display devices.
On the other hand, topological point and line defects in the liquid
crystalline order, which occur as a consequence of the
broken rotational symmetry of the isotropic space
\cite{itapdb:Kleman1983,itapdb:Mermin1979,itapdb:Trebin1982,itapdb:Kurik1988,itapdb:Chaikin1995},
attract a lot of attention since they are fascinating by themselves.
They, furthermore, influence the static and dynamic properties of the
system under investigation.
Conventional displays, like twisted nematic \cite{itapdb:Schadt1971} or 
surface-stabilized ferroelectric liquid crystal cells 
\cite{itapdb:Clark1980}, 
possess a simple geometry, where the liquid crystal
is enclosed between two parallel glass plates. With the introduction
of polymer-dispersed liquid crystals as electrically controllable 
light shutters \cite{itapdb:Doane1986,itapdb:Drzaic1995}, an extensive
study of liquid crystals confined to complex geometries, like
drops in a polymer matrix or a random porous network in silica aerogel,
was initiated \cite{itapdb:Drzaic1995,itapdb:Crawford1996}.

%On the other hand, topological point and line defects in the
%orientational order of the molecules occur as a consequence of the 
%broken rotational symmetry of the isotropic space
%\cite{itapdb:Kleman1983,itapdb:Mermin1979,itapdb:Trebin1982,itapdb:Kurik1988,itapdb:Chaikin1995}.
%These defects are fascinating in itself to study, and, 
%furthermore, influence the static and the dynamic properties of the
%system under investigation.

This article presents a numerical investigation of the interesting
inverse problem, which is posed by particles suspended in a nematic solvent.
In a nematic liquid crystal the molecules align on 
average along a unit vector $\bbox{n}$, called director. 
The energetic ground
state is a uniform director field throughout space. Due to the
anchoring of the molecules on the surface of a particle the 
surrounding director field is elastically distorted. 
In addition to conventional
van der Waals forces, screened Coulomb and steric interactions 
\cite{itapdb:Russel1995}, the elastic deformation of the
director field mediates a further interaction between the
particles. Point defects give rise to a short-range repulsion.

Already in 1970, Brochard and de Gennes studied a suspension of
magnetic grains in a nematic liquid crystal and determined the
director field far away from a particle \cite{itapdb:Brochard1970}. 
A bistable liquid crystal display was introduced based on a dispersion of
agglomerations of silica spheres in a nematic host 
\cite{itapdb:Eidenschink1991,itapdb:Kreuzer1992,itapdb:Glushchenko1997}.
Chains and clusters were observed in the dispersion of latex particles 
in a lyotropic liquid crystal 
\cite{itapdb:Poulin1994,itapdb:Raghunathan1996,itapdb:Raghunathan1996a}. 
The radii of the particles were 60 and 120$\,$nm. Therefore, details
of the director field could not be resolved under the polarizing
microscope.
Terentjev {\em et al.\/} investigated the director field around a sphere
by both analytical and numerical methods
\cite{itapdb:Terentjev1995,itapdb:Kuksenok1996}. 
For homeotropic boundary conditions of the director they
considered the Saturn-ring configuration where the sphere is
surrounded by a $-1/2$ disclination ring. The long-range quadrupolar 
interaction of such objects was determined by Ramaswamy {\em et al.}
\cite{itapdb:Ramaswamy1996} and Ruhwandl and Terentjev 
\cite{itapdb:Ruhwandl1997}. 
Even two parallel plates, immersed into the isotropic phase close to the
isotropic-nematic phase transition, interact as a result of a
surface-induced nematic order \cite{itapdb:Borstnik1997}.

\subsection{Inverted Nematic Emulsions}

The present article addresses recent work on inverted nematic emulsions
where surfactant-coated water droplets are dispersed in a nematic
solvent \cite{itapdb:Poulin1997,itapdb:Poulin1998}.
The advantage of such a system is that the particles are
easily observable by polarizing microscopy since the size of the
water droplets is of the order of a micron. Furthermore, the anchoring of
the liquid crystal molecules on the surfaces of the droplets
are controllable via the surfactant. Understanding the suspension of
such large particles will, without any doubt, help to clarify the
observations of dispersions of smaller objects. The most striking
feature, from a theoretical point of view, is that inverted emulsions 
provide an ideal laboratory for the investigation of topological
defects. In inverted emulsions point defects, also called hedgehogs,
occur. They carry a topological charge $q$ specifying the
number of times the unit sphere is wrapped by the director $\bbox{n}$
on any surface enclosing the defect core. Since all properties of the
nematic phase are invariant under the inversion
$\bbox{n} \rightarrow -\bbox{n}$, $q$ is always positive,
and two hedgehogs with respective charges $q_{1}$ and $q_{2}$ ``add
up'' to a total topological charge of either $q_{1} + q_{2}$ or 
$|q_{1} - q_{2}|$ \cite{itapdb:Kurik1988}.

\begin{figure}
\centerline{\psfig{figure=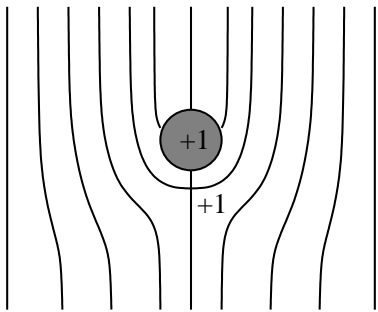,width=5cm}}

\vspace{.5cm}

\caption[]{The water droplet and its companion hyperbolic hedgehog form a
  dipole. Both the droplet and the hedgehog carry a topological charge
 $+1$, and the total charge of the dipole is $0=|1-1|$.
 }

\vspace{.3cm}

\label{f.drophedge}
\end{figure}

When a water droplet with homeotropic, {\em i.e.}, 
perpendicular anchoring of the
director at its surface is placed into a uniformly aligned nematic liquid
crystal, a hyperbolic hedgehog is nucleated (see Fig. \ref{f.drophedge}).
The droplet and its tightly bound companion defect provide a key unit 
to understand inverted nematic emulsions
\cite{itapdb:Meyer1972,itapdb:Poulin1997,itapdb:Poulin1998,itapdb:Lubensky1998}.
We will call it a dipole for short because of its dipolar symmetry. 
Both the water droplet and the hyperbolic hedgehog carry a topological charge 
$+1$ which ``add up'' to the total charge 0 of the dipole.
A phenomenological theory predicts that the dipole couples to a
strong splay deformation in the director field 
\cite{itapdb:Poulin1997,itapdb:Lubensky1998}.
Two parallel dipoles are attracted via a long-range dipolar interaction
which explains the observed chaining of water droplets
\cite{itapdb:Poulin1997,itapdb:Poulin1998,itapdb:Lubensky1998}. A 
quantitative confirmation of the dipolar interaction was presented recently 
\cite{itapdb:Poulin1997a}.

In our numerical investigation we consider multiple nematic emulsions
where the nematic host fluid containing the water droplets is confined to
large nematic drops, which, in turn, are surrounded by water. 
The water \linebreak

\begin{figure}
\centerline{\psfig{figure=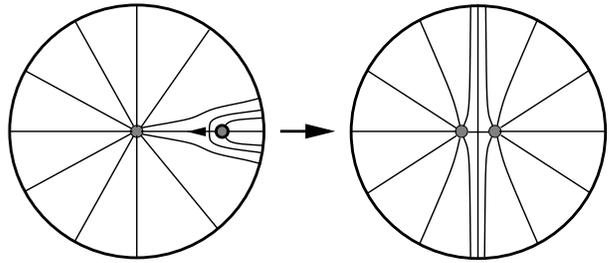,width=8cm}}

\vspace*{.3cm}

\caption[]{Scenario to explain the chaining of water droplets in a large
  nematic drop. The right water droplet and its companion hyperbolic 
  hedgehog form a dipole, 
  which is attracted by the strong splay deformation around the
  droplet in the center (left picture). The dipole moves towards the
  center until at short distances the repulsion mediated by
  the point defect sets in (right picture).}
\label{f.chain}
\end{figure}
\noindent
droplets also form chains with a
hyperbolic hedgehog situated between two droplets 
(see Fig.\ \ref{f.chain}, right)
\cite{itapdb:Poulin1997,itapdb:Poulin1998}. In the experiment it was
found that the 
distance $d$ of the point defect from the surface of a water droplet scales
with the radius $r$ of the droplet like $d \approx 0.3 \, r$
\cite{itapdb:Poulin1997,itapdb:Poulin1998}. In the
following we will call this relation the scaling law. A possible
explanation for the chaining, as in a uniformly aligned sample, could
be the presence of a dipole. One water droplet fits perfectly into the
center of a large nematic drop, which has a total topological charge $+1$.
Any additional water droplet has to be accompanied by a hyperbolic
hedgehog in order not to change the total charge. If the dipole forms
(see Fig. \ref{f.chain}, left), it is attracted by the strong splay
deformation in the center \cite{itapdb:Lubensky1998}, until a short-range 
repulsion mediated by the defect sets in 
\cite{itapdb:Poulin1997,itapdb:Poulin1998}.

\subsection{Main Results}

By our numerical investigations we confirm the scaling law, for which
we find $d = (0.325 \pm 0.025) \, r$, and discuss the influence of the outer
boundary of the large drop. We, furthermore, show that in nematic drops the 
dipole also exists, confirming the scenario of Fig.\ \ref{f.chain}.
When the two water droplets in the right picture of Fig.\ \ref{f.chain}
are moved apart symmetrically about the center of the large drop, 
the dipole forms via a second order phase transition.
We also identify the dipole in a bipolar configuration
which occurs for planar boundary conditions at the outer surface
of the nematic drop. Two boojums, {\em i.e.}, surface defects appear
\cite{itapdb:Mermin1977,itapdb:Candau1973,itapdb:Kurik1982}, and the
dipole is attracted by the strong splay deformation in the vicinty of
one of them \cite{itapdb:Poulin1997,itapdb:Poulin1998,itapdb:Lubensky1998}. 
Besides the dipole we find another stable configuration in this geometry,
where the hyperbolic hedgehog sits close to one of the boojums. Finally,
we show that water droplets can repel each other without a hyperbolic
defect placed between them.

To minimize the Oseen-Z\"ocher-Frank free energy \cite{itapdb:deGennes1993}
for different
positions of the water droplets we employ the method of finite elements 
\cite{itapdb:Twizell1984},
which is most suitable for  \linebreak
\newpage
\begin{figure}
\centerline{\psfig{figure=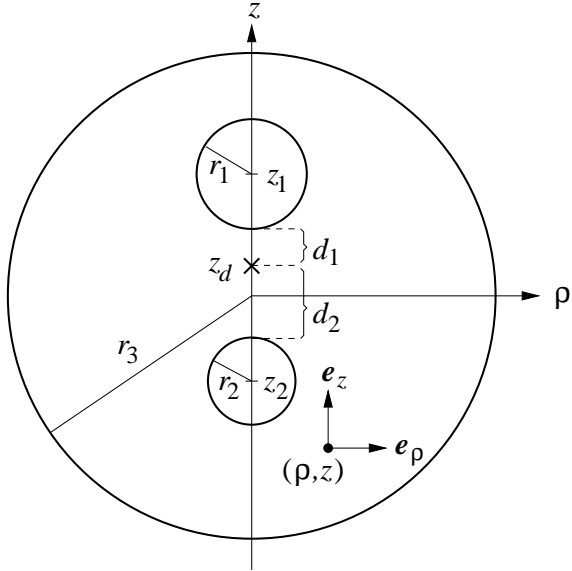,width=7.5cm}}

\vspace{.3cm}

\caption[]{Geometry parameters for two water droplets with respective radii
$r_{1}$ and $r_{2}$ in a large nematic drop with radius
$r_{3}$. The system is axially symmetric about the $z$ axis, and
cylindrical coordinates $\rho,z$ are used. The coordinates $z_{1}$,
$z_{2}$, and $z_{d}$ are the respective positions of the two droplets
and the hyperbolic hedgehog. The two distances of the hedgehog from
the surfaces of the droplets are $d_{1}$ and $d_{2}$.}
\label{f.geo}
\end{figure}
\noindent
non-trivial geometries. Since we are not aware that this
method has been used before in liquid crystal research,
we provide a more detailed description of it in Sec.\ \ref{sec.num}
and in the Appendix. We hope to stimulate further applications of
the method of finite elements to liquid crystals in complex
geometries. Before we discuss our results in Sec.\ \ref{sec.disc}
we define the geometries and present numerical details in 
Sec.\ \ref{sec.num}.

\section{Geometries and Numerical Methods} \label{sec.num}
%\section{Defining the Problem and Numerical Method} \label{sec.num}
%\section{Numerical Realization}

In our article we numerically investigate two particular geometries 
of axial symmetry. The first problem is defined in
Fig.\ \ref{f.geo}. We consider two spherical water droplets with
respective radii $r_{1}$ and $r_{2}$ in a large nematic drop with
radius $r_{3}$.
The whole system possesses axial symmetry, so that
the water droplets and the hyperbolic hedgehog, indicated by a cross,
are located always on the $z$ axis. We employ a cylindrical coordinate system.
The coordinates $z_{1}$, $z_{2}$, and $z_{d}$ denote, respectively,
the positions of the centers of the droplets and of the hyperbolic
hedgehog on the $z$ axis.
The distances of the hedgehog from the surfaces of the two water droplets
are, respectively, $d_{1}$ and $d_{2}$. Then, the quantity $d_{1} + d_{2}$
means the distance of the two small spheres, and the point defect
is situated in the middle between them if $d_{1}=d_{2}$.
We, furthermore, restrict the nematic director to the $(\rho,z)$ plane,
which means that we do not allow for twist deformations. The director
is expressed in the local coordinate basis $(\bbox{e}_{\rho},\bbox{e}_{z})$,
\begin{equation}
\label{1}
\bbox{n}(\rho,z) = \sin\Theta(\rho,z) \bbox{e}_{\rho} + 
                   \cos\Theta(\rho,z) \bbox{e}_{z} \enspace,
\end{equation}
where we introduced the {\em tilt angle\/} $\Theta$. It is always 
restricted to
the range $[-\pi/2,\pi/2]$ to ensure the $\bbox{n} \rightarrow -\bbox{n}$
symmetry of the nematic phase. At all the boundaries we assume
a rigid homeotropic anchoring of the director, which allows us to omit
any surface term in the free energy.

In the second problem we have only one water droplet inside a large
nematic drop. We use the same coordinates and lengths as described in
Fig. \ref{f.geo}, but omit the second droplet. The anchoring of
the director at the outer surface of the large nematic sphere is 
rigid planar. At the surface of the small sphere we again choose a 
homeotropic boundary condition.

We write the Oseen-Z\"ocher-Frank free energy density 
\cite{itapdb:deGennes1993} for the director field (\ref{1})
in units of $K_{33} /a^{2}$, where $K_{33}$ is the bend elastic constant
and $a$ the characteristic length scale of our system, typically
several microns:
\begin{eqnarray}
\overline{f}(\Theta) & = & \frac{f(\Theta)}{K_{33}/a^{2}} \nonumber\\
 & = & \frac{\overline{K}_{11}}{2} \, \left( \frac{\sin\Theta}{\rho} 
+ \Theta_{\rho}\,\cos\Theta - \Theta_z\,\sin\Theta \right)^2 \nonumber\\
\label{2}
 & & + \, \frac{1}{2} \,
\left(\Theta_{z}\,\cos\Theta + \Theta_{\rho}\,\sin\Theta \right)^2 \enspace.
\end{eqnarray}
All lengths are given relative to $a$, and $\overline{K}_{11} =
 K_{11}/K_{33}$ is the scaled splay constant. The indices $\rho$ and $z$
indicate, respectively, partial derivatives $\partial/\partial \rho$
and $\partial/\partial z$.

\begin{figure}
\centerline{\psfig{figure=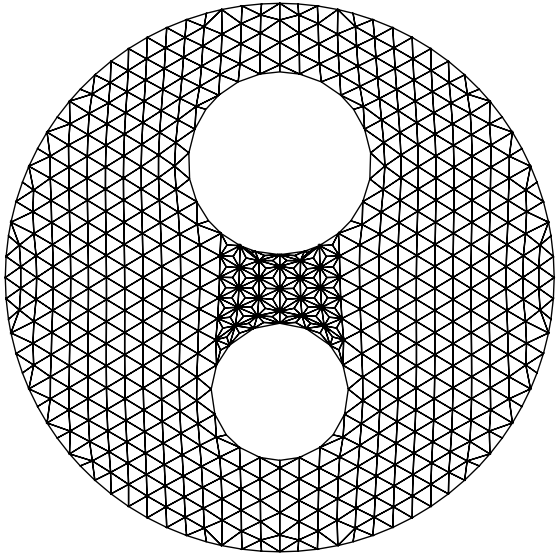,width=7.5cm}}

\vspace{.5cm}

\caption[]{Triangulation of the integration area (lattice constant: $b
  = 0.495$). Between the small spheres a refined net of triangles is chosen.}
\label{f.net}
\end{figure}

In order to find the equilibrium director field
%Frank free energy
the reduced free energy in units  of $K_{33} a$,
\begin{equation}
\overline{F} = \int d^{3}x  \, \overline{f}(\Theta) \enspace,
\end{equation}
is minimized on a grid. Because of the nontrivial geometry of our
problem, we decided to employ the method of finite elements 
\cite{itapdb:Twizell1984},
where the integration area is covered with triangles. We note, that
for simpler geometries the method of finite differences is usually used,
where the grid is defined by the coordinate lines. We construct
a net of triangles by covering our integration area with a hexagonal
lattice with lattice constant $b$.
Vertices of triangles that only partially belong to the
integration area are moved onto the boundary along the radial
direction of the appropriate sphere. As a result, extremely obtuse
triangles occur close to the boundary. We use a relaxation
mechanism to smooth out these irregularities. The final triangulation 
is shown in Fig. \ref{f.net}. In the area between the small spheres, 
where the hyperbolic hedgehog is situated, the grid is further subdivided to
account for the strong director deformations close to the point defect.

The reduced free energy $\overline{F}$ in units of $K_{33} a$ is written
as a function of the tilt angles
$\Theta_{i}$ ($i=1, \dots ,n$) at the $n$ \linebreak
\noindent
vertices of the grid:
$\overline{F} = \overline{F}[\bbox{\Theta}]$, where
$\bbox{\Theta} = (\Theta_{1}, \ldots, \Theta_{n})$.
For further explanations how $\overline{F}[\bbox{\Theta}]$ is
constructed within the method of finite elements we refer the reader to
the Appendix.
To find a minimum of the free energy we start with a configuration
that already possesses the hyperbolic point defect at $z_{d}$ and let it 
relax via the standard Newton-Gauss-Seidel method \cite{itapdb:Press1992}:
\begin{equation}
\Theta_{i}^{\text{new}} = \Theta_{i}^{\text{old}} -
\frac{\partial \overline{F}[\bbox{\Theta}]}{\partial \Theta_{i}}
\left/ 
\frac{\partial^{2} \overline{F}[\bbox{\Theta}]}{\partial \Theta_{i}^{2}}
\right. \enspace.
\end{equation}
The derivatives $\overline{F}$
with respect to $\Theta_{i}$ are performed numerically.
Although the point defect is not fixed to its original position
during the relaxation it never moves 
%completely
into the global minimum, but rather stays where we place it.
We use this fact to plot energy curves versus the position of
the point defect.
Other starting configurations, {\em e.g.}, a uniform director field
throughout the integration area or randomly oriented directors on the grid,
do not produce the hyperbolic hedgehog, but instead relax into 
a configuration with a higher energy, which we
briefly investigate in subsection \ref{subsec.rep}.

Integrating the free energy density over one 
triangle yields a line energy, {\em i.e.}, an energy per unit length.
As a rough estimate for its upper limit we
introduce the line tension $f_{l} = (\overline{K}_{11} + 1) /2$ of the 
isotropic core of a disclination \cite{itapdb:deGennes1993}. Whenever the 
numerically calculated local line energy is larger than $f_{l}$, 
we replace it by $f_{l}$.
As a result, the hyperbolic point defect is stabilized against
opening up to a disclination ring.

All our calculations are preformed for the nematic liquid crystal
pentylcyanobiphenyl (5CB), for which the experiments were done
\cite{itapdb:Poulin1997,itapdb:Poulin1998}.
It has a bend elastic constant
$K_{33} = 0.53 \times 10^{-6} \text{dyn}$ and a scaled splay elastic constant
$\overline{K}_{11} = K_{11}/K_{33} = 0.79$.
The experimental ratio $r_{3}/r_{1/2}$ of the radii of the
large and small drops is in the range $10-50$ 
\cite{itapdb:Poulin1997,itapdb:Poulin1998}.
The difficulty is that we want to investigate
details of the director field close to the small spheres which requires
a fine triangulation on the length scale given by $r_{1/2}$. To keep
the computing time to a reasonable value we choose 
the following lengths: 
$r_{3} = 7$, $r_{1/2} = 0.5 \ldots 2$, and $b=0.195$ for the
lattice constant of the grid. 
In addition, we normally use one step of grid refinement
between the small spheres (geometry 1) or 
between the small sphere and the south pole of the large nematic drop
(geometry 2). With such parameters we obtain a lattice with 
2200-2500 vertices.

\section{Results and Discussion} \label{sec.disc}

In this section we discuss the results from our numerical investigation.
We will mainly address the following questions:\\
1. Can we confirm the scaling law $d_{1/2} \approx 0.3 \, r_{1/2}$,
which was observed in experiment, by varying the different lengths
in our geometry?\\
2. Does the dipole formed by a water droplet and its companion
hyperbolic defect, which is well-established for a director field 
uniformly aligned at infinity, also have a meaning
in more complex geometries?\\
3. Is the hyperbolic hedgehog necessary to mediate a repulsion between
the water droplets? To answer this question we will investigate a 
configuration with disclination rings around the small spheres.

\subsection{Scaling Law}

\begin{figure}
\centerline{\psfig{figure=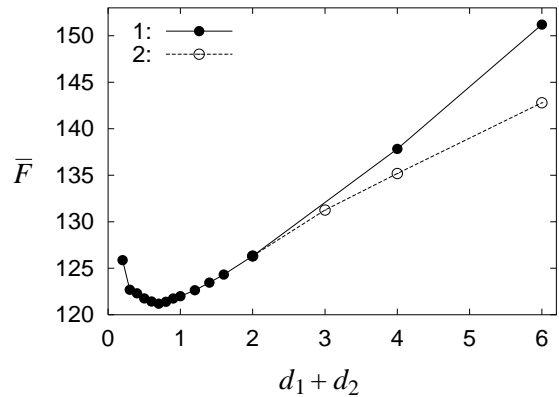,width=7.5cm}}

\vspace{.5cm}

\caption[]{The free energy $\overline{F}$ as a function of the distance 
$d_{1}+d_{2}$ between the small spheres which are placed symmetrically
about $z=0$ ($r_{1}=r_{2}=1$). 
Curve 1: $z_{d} =0$, curve 2: position $z_{d}$ of the defect can relax 
along the $z$ axis.}
\label{f.curve1}
\end{figure}

We start with the first question. In Fig.\  \ref{f.curve1} we plot the reduced
free energy $\overline{F}$ as a function of the distance \linebreak
\noindent
$d_{1}+d_{2}$
between the surfaces of the small spheres, which are placed
symmetrically about the center, {\em i.e.}, $z_{2} = -z_{1}$. 
Their radii are $r_{1}=r_{2}=1$. Curve 1 shows a clear
minimum at $d_{1} + d_{2} \approx 0.7$, the defect stays in the
middle between the two spheres at $z_{d} = 0$. In curve 2 we move the
defect along the $z$ axis and plot the minimum free energy for each
fixed distance $d_{1}+d_{2}$. It is obvious that beyond 
$d_{1}+d_{2} = 2$ the defect moves to one of the small spheres. We
will investigate this result in more detail in the following subsection.

\begin{figure}
\centerline{\psfig{figure=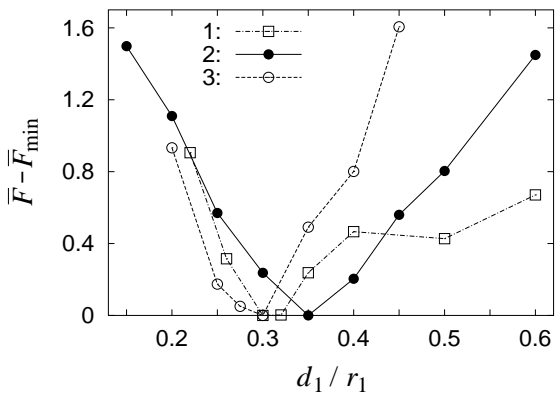,width=7.5cm}}

\vspace{.5cm}

\caption[]{The free energy $\overline{F} - \overline{F}_{\mathrm{min}}$ 
as a function of $d_{1}/r_{1} = d_{2} / r_{2}$. The small spheres are 
placed symmetrically about $z=0$. 
Curve 1: $r_{1}=r_{2}=0.5$, curve 2: $r_{1} = r_{2} = 1$,
and curve 3: $r_{1}=r_{2}=2$.}
\label{f.curve2}
\end{figure}

In Fig.\ \ref{f.curve2} we take three different radii for the small
spheres, $r_{1} = r_{2} = 0.5,1,2$, and plot the free energy
versus $d_{1}/r_{1}$ close to the minimum. Recall that $d_{1}$ is the 
distance of the hedgehog from the surface of sphere 1. 
Since for such small distances $d_{1}+d_{2}$
the defect always stays at $z_{d}=0$, {\em i.e.}, in the middle
between the two spheres, we have $d_{1}/r_{1} = d_{2}/r_{2}$.
The quantity $\overline{F}_{\mathrm{min}}$ refers to the minimum free
energy of each curve.
For each of the three radii we obtain an energetically preferred
distance $d_{1}/r_{1}$ in the range of [0.3,0.35], which agrees well
with the experimental value of 0.3. Why does a scaling law
of the form $d_{1/2} = (0.325 \pm 0.025) \, r_{1/2}$ occur? If the small
spheres are far away from the surface of the large nematic drop,
its finite radius $r_{3}$ should hardly influence the distances 
$d_{1}$ and $d_{2}$. Then, the only length scale in the system is 
$r_{1}=r_{2}$, and we expect $d_{1/2} \propto r_{1/2}$. 
However, in Fig.\ \ref{f.curve2} the influence from the boundary of
the large sphere is already visible.
Let us take curve 2 for spheres with radii $r_{1/2} = 1$ as
a reference. It looks pretty symmetric around $d_{1}/r_{1} = 0.35$.
%Let us take curve 2 ($r_{1} = r_{2} = 1$: $\bullet$), which looks
%pretty symmetric around $d_{1}/r_{1} = 0.35$, as a reference.
The slope of the right part of curve 3, which corresponds to larger
spheres of radii $r_{1/2} = 2$, is steeper than in curve 2.
%($r_{1} = r_{2} = 2$: $\circ$) is steeper than in curve 2,
Also, the location of the minimum clearly tends to values smaller than
0.3. We conclude that the small spheres are already so large
that they are strongly repelled by the boundary of the nematic drop.
On the other hand, the slope of the 
right part of curve 1, which was
calculated for spheres of \linebreak
\noindent
radii $r_{1/2} = 0.5$, is less steep than in curve 2. This leads
to the conclusion that the boundary of
the nematic drop has only a minor influence on such small spheres.
%In curve 1 ($r_{1} = r_{2} =0.5$: $\Box$), however, the slope of the
%right part is less steep than in curve 2. Since for $r_{1/2}=0.5$ the
%boundary is farther away than for $r_{1/2}=1$, its presence has only little
%influence on the water droplets
%The boundary is farther away than in
%case 2 ($r_{1}=r_{2}=1$) and has only little influence.

When we move the two spheres with radii $r_{1/2}=1$ together in the 
same direction along the $z$ axis, the defect always stays in the middle
between the droplets and obeys the scaling law. We have tested its 
validity within the range [0,3] for the defect position $z_{d}$. 
Of course, the absolut minimum of the free energy occurs in the 
symmetric position of the two droplets, $z_{2} = -z_{1}$. 

We further check the scaling law for $r_{1} \ne r_{2}$.
We investigate two cases. When we choose $r_{1} = 2$ and $r_{2} = 0.6$, 
we obtain $d_{1/2} \approx 0.3 \, r_{1/2}$.
In the second case, $r_{1}=2$ and $r_{2} = 1$, 
we find $d_{1} \approx 0.37 \, r_{1}$ and $d_{2} \approx 0.3 \, r_{2}$.
As observed in the experiment the defect sits always closer to
the smaller sphere.
There is no strong deviation from the scaling law 
$d_{1/2} = (0.325 \pm 0.025) \, r_{1/2}$, although we would allow for it,
since $r_{1} \ne r_{2}$.

%The first one is for
%$r_{1} = 2$ and $r_{2} = 0.6$ and results in $d_{1/2} \approx 0.3
%r_{1/2}$. In the second case, $r_{1}=2$ and $r_{2} = 1$, we obtain
%$d_{1} \approx 0.37 r_{1}$ and $d_{2} \approx 0.3 r_{2}$.
%Clearly, the defect sits always closer to the smallest sphere.

\subsection{Identification of the Dipole}

\begin{figure}
\centerline{\psfig{figure=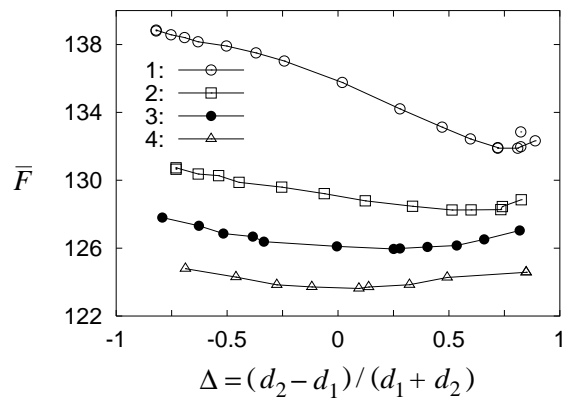,width=7.5cm}}

\vspace{.5cm}

\caption[]{The free energy $\overline{F}$ as a function of\\
$\Delta = (d_{2} - d_{1}) / (d_{1} + d_{2})$. Sphere 2 is placed at
$z_{2} = 0$. The position $z_{1}$ of sphere 1 is the parameter. 
Curve 1: $z_{1}=5$, curve 2: $z_{1}=4$, curve 3: $z_{1}=3.5$,
and curve 4: $z_{1}=3$. The radii are $r_{1}=r_{2}=1$.}
\label{f.curve3a}
\end{figure}

The second question is if the dipole formed by one water droplet
and a companion hyperbolic point defect has a meaning in our geometry.
To answer this question we place sphere 2 with radius $r_{2}=1$
in the center of the nematic drop at $z_{2}=0$. Then, we determine the
energetically preferred position of the point defect for different 
locations $z_{1}$ of sphere 1 ($r_{1}=1$). The position of the hedgehog 
is indicated by $\Delta = (d_{2} - d_{1}) / (d_{1} + d_{2})$. If the defect 
is located in the middle between the two spheres, 
$\Delta$ is zero since $d_{1}=d_{2}$.
On the other hand,
if it sits at the surface of sphere 1, $\Delta$ is one since $d_{1} = 0$.
In Fig. \ref{f.curve3a} we plot the free energy $\overline{F}$ 
versus $\Delta$. In curve 1, where the small
spheres are farthest apart from each other ($z_{1} = 5$), \linebreak
\noindent
we clearly find the defect close to sphere 1. 
This verifies that the dipole is existing. It is
thermally stable, since a rough estimate of the thermally induced
mean displacement of the defect yields 0.01 
\cite{itapdb:Lubensky1998}. When sphere 1
is approaching the center (curve 2: $z_{1} = 4$ and curve 3:
$z_{1} = 3.5$), the defect moves away
from the droplets until it nearly reaches the middle between both spheres
(curve 4: $z_{1} = 3$). This means, the dipole vanishes gradually
until the hyperbolic hedgehog is shared by both water droplets.

\begin{figure}
\centerline{\psfig{figure=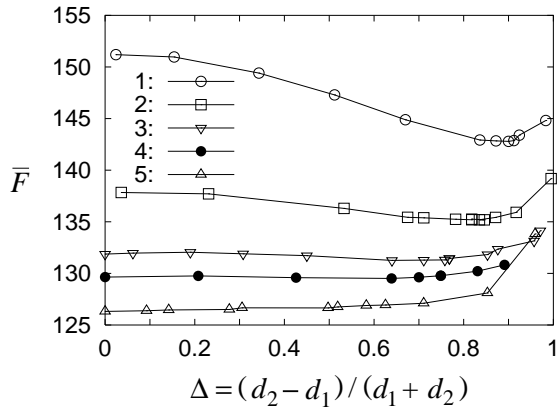,width=7.5cm}}

\vspace{.5cm}

\caption[]{The free energy $\overline{F}$ as a function of\\
$\Delta = (d_{2} - d_{1}) / (d_{1} + d_{2})$. The small spheres are
placed symmetrically about $z = 0$.
Curve 1: $z_{1}=-z_{2}=4$, curve 2: $z_{1}=-z_{2}=3$, curve 3: 
$z_{1}=-z_{2}=2.5$, curve 4: $z_{1}=-z_{2}=2.3$, curve 5: $z_{1}=-z_{2}=2$. 
The radii are $r_{1}=r_{2}=1$.}
\label{f.curve3}
\end{figure}

An interesting situation occurs when sphere 1 and 2 are placed
symmetrically about $z=0$. Then, the defect has two equivalent
positions on the positive and negative part of the $z$ axis.
In Fig. \ref{f.curve3} we plot again the free
energy $\overline{F}$ versus the position $\Delta$ of the defect.
>From curve 1 to 3 ($z_{1}=z_{2}=4,\,3,\,2.5$) the minimum in 
$\overline{F}$ becomes broader and more shallow. 
The defect moves closer towards the center
until at around $z_{1}=-z_{2}=2.3$ (curve 4) it reaches $\Delta = 0$.
This is reminiscent to a symmetry-breaking second order phase
transition \cite{itapdb:Chaikin1995,itapdb:Landau1984f}
which occurs when, in the course of moving the
water droplets apart, the dipole starts to form. We
take $\Delta$ as an order parameter, where $\Delta = 0$ and
$\Delta \ne 0$ describe, respectively, the high- and the low-symmetry
phase. A Landau expansion of the free energy yields
\begin{equation}
\overline{F}(\Delta) = \overline{F}_{0}(z_{1}) +
a_{0} [2.3 - z_{1}] \Delta^{2}  + c (z_{1}) \Delta^{4} \enspace,
\end{equation}
where $z_{1} = -z_{2}$ plays the role of the temperature.
Odd powers in $\Delta$ are not allowed because of the required symmetry
$\overline{F}(\Delta)  = \overline{F}(-\Delta)$. This free energy
qualitatively describes the curves in Fig. \ref{f.curve3}. 
It should be possible to observe such a second order phase transition
with a method introduced recently by Poulin {\em et al.\/} 
\cite{itapdb:Poulin1997a}
to measure dipolar forces in inverted nematic emulsion. When the two small
droplets are filled with a magnetorheological fluid
\linebreak

\begin{figure}
\centerline{\psfig{figure=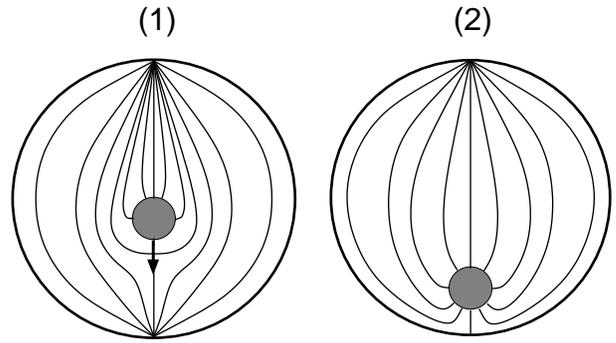,width=8cm}}

\vspace{.5cm}

\caption[]{Planar boundary conditions on the outer surface of the large
  sphere create boojums, {\em i.e.}, 
  surface defects at the north and the south
  pole. A water droplet with homeotropic boundary conditions nucleates
  a hyperbolic hedgehog. 
  Two configurations exist that are either stable or metastable
  depending on the position of the water droplet;
  (1) the dipole, (2) the hyperbolic hedgehog sitting at the surface.}
\label{f.dipole}
\end{figure}
\noindent
instead of pure water, a small magnetic field of about 
100 G, applied perpendicular to the $z$ axis,
induces two parallel magnetic dipoles. Since they repel each other,
the two droplets are forced apart. When the magnetic field is switched off,
the two droplets move towards each other to reach the equilibrium
distance. In the course of this process the phase transition for
the dipole should be observable.

\subsection{The Dipole in a Bipolar Configuration} \label{subsec.bipo}

It is possible to change the anchoring of the director on the outer
surface of the large nematic drop from homeotropic to planar
by adding some amount of glycerol to the surrounding water phase
\cite{itapdb:Poulin1997}.
Then, the bipolar configuration for the director field appears 
\cite{itapdb:Candau1973,itapdb:Kurik1982}, where two boojums 
\cite{itapdb:Mermin1977}, 
{\em i.e.}, surface defects of charge 1 are situated at the north and south 
pole of the large nematic drop
(see configuration (1) in Fig.\ \ref{f.dipole}).
The topological point charge of the interior of the nematic drop is zero, 
and every small water droplet with homeotropic boundary condition
has to be accompanied by a hyperbolic hedgehog. In the experiment the 
hedgehog sits close to the water droplet, {\em i.e.}, the
dipole exists and it is attracted by the strong splay deformation close to
the south pole \cite{itapdb:Poulin1997}, as predicted by the phenomenological
theory \cite{itapdb:Poulin1997,itapdb:Lubensky1998}.

\begin{figure}
\centerline{\psfig{figure=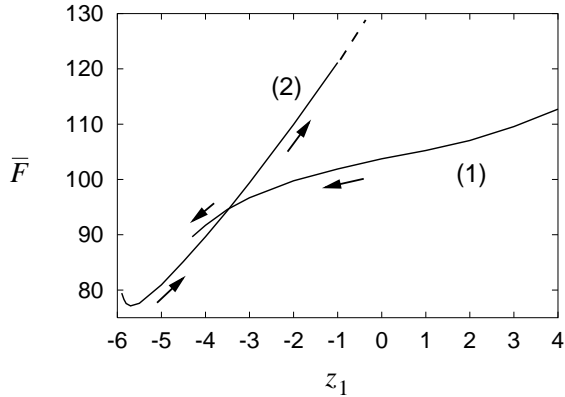,width=7.5cm}}

\vspace{.5cm}

\caption[]{The free energy $\overline{F}$ as a function of the position
  $z_{1}$ of the water droplet for the configurations (1) and (2). For
  $z_{1} > -3.5$, (1) is stable, and (2) is metastable. The situation is
  reversed for $-4.3 < z_{1} < -3.5$. Configuration (1) looses its 
  metastability at $z_{1} = -4.3$.}
\label{f.curve4}
\end{figure}

A numerical analysis of the free energy $\overline{F}$
is in agreement with experimental observations but also reveals some
interesting details which have to be confirmed. In Fig.\ \ref{f.curve4}
we plot $\overline{F}$ as a function of the position $z_{1}$ of the
small water droplet with radius $r_{1} = 1$. The diagram consists of 
curves (1) and (2), which correspond, respectively, to configurations (1) 
and (2) in Fig.\ \ref{f.dipole}.
The free energy possesses a minimum at around $z_{1} = -5.7$.
The director field
assumes configuration (2), where
the hyperbolic hedgehog 
is situated at the surface of the nematic drop. \linebreak
\noindent
Moving the water droplet closer to the surface induces a repulsion due
to the strong director deformations around the point defect.
When the water droplet is placed far away from the south pole, 
{\em i.e.},
at large $z_{1}$, the dipole of configuration (1) forms and represents
the absolute stable director field. At $z_{1} = -3.5$ the dipole
becomes metastable but the system does not assume configuration (2) 
since the energy barrier, the system has to overcome by thermal activation,
is much too high. By numerically calculating the free energy for
different positions of the hedgehog
we have, {\em e.g.}, at $z_{1}=-4.0$ determined an 
energy barrier of $K_{33}a \approx 1000 \, k_{\mathrm{B}}T$, where 
$k_{\mathrm{B}}$ is the Boltzmann constant, $T$ the room temperature,
and $a \approx 1\,\mu \text{m}$. At $z_{1} = -4.3$ the dipole even
looses its metastability, the
hyperbolic defect jumps to the surface at the south pole and the water
droplet follows until it reaches its energetically preferred position.
On the other hand, if it were possible to move the water droplet away
from the south pole, the hyperbolic hedgehog would stay at the
surface, since configuration (2) is always metastable for $z_{1} \ge
-3.5$. The energy barrier for a transition to the dipole is again
at least 1000 $k_{\mathrm{B}}T$. We have also investigated the
distance $d_{1}$ of the defect from the surface of the water droplet.
For $z_{1} \in [-2,4]$, $d_{1}$ fluctuates between $0.3$ and
$0.35$. For $z_{1} < -2$, it increases up to $0.5$ at $z_{1} =
-4.3$, where the dipole looses its metastability.

\subsection{Repulsion without Defect} \label{subsec.rep}

\begin{figure}
\centerline{\psfig{figure=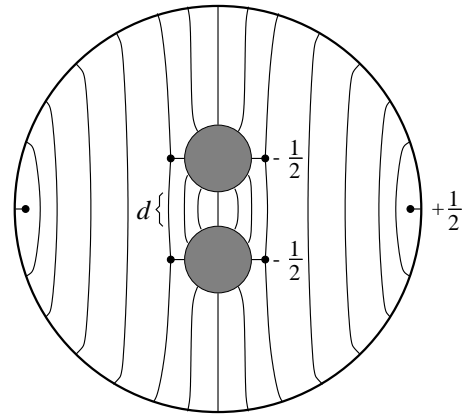,width=6cm}}

\vspace{.5cm}

\caption[]{An alternative metastable configuration. Both droplets
  are surrounded by a $-1/2$ disclination ring which compensates the
   topological charge $+1$ of each droplet. An additional $+1/2$
   disclination ring close to the surface of the nematic drop
   satisfies the total topological charge $+1$.}
\label{f.saturn}
\end{figure}

We return to the first geometry with two water droplets and homeotropic
boundary conditions at all the surfaces. 
When we take either a uniform director
field or randomly oriented directors as a starting configuration,
our system always relaxes into the configuration sketched in 
Fig.\ \ref{f.saturn}. \linebreak
\noindent
Both water droplets are surrounded in their
equatorial plane by a $-1/2$ disclination ring which compensates
the point charge $+1$ each droplet carries with it 
\cite{itapdb:Terentjev1995,itapdb:Kuksenok1996}. To obtain the
total point charge $+1$ of the nematic drop there has to be an
additional topological defect with a point charge $+1$. In the numerically
relaxed director field we find a $+1/2$ disclination ring close to
the outer surface. This configuration has a higher energy than the
one with the hyperbolic hedgehog. It is only metastable.
Since a transition to the stable configuration needs a complete 
rearrangement of the director field, the energy barrier is certainly
larger than $K_{33}a \approx 1000 \,k_{\mathrm{B}}T$. We, therefore,
expect the configuration of Fig.\ \ref{f.saturn} to be stable against
thermal fluctuations. It would be interesting to search for it in the
experiment.

\begin{figure}
\centerline{\psfig{figure=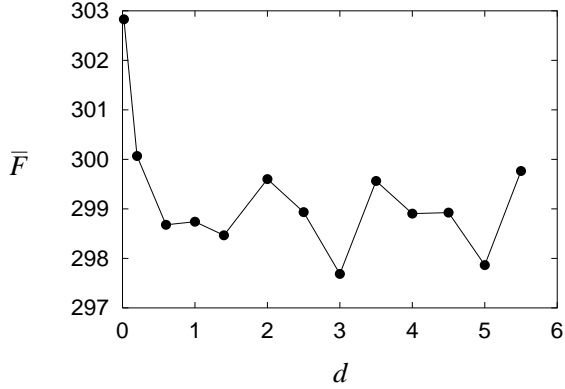,width=7.5cm}}

\vspace{.5cm}

\caption[]{The free energy $\overline{F}$ as a function of the
  distance $d$ of the droplets. A repulsion for $d < 0.6$ is clearly
  visible.}
\label{f.curve5}
\end{figure}

We use the configuration to demonstrate that even without the hyperbolic
hedgehog the two water droplets experience some repulsion when they
come close to each other. In Fig.\ \ref{f.curve5} we plot the free
energy $\overline{F}$ versus the distance $d$ of the two spheres.
For large $d$ the free energy oscillates which we attribute to
numerical artifacts. For decreasing $d$ the free energy clearly 
increases, and the water droplets repel each other due to the strong 
deformation of the director field lines connecting the two droplets.

\acknowledgements
We thank A. Emerson, 
A. Kilian, A. R\"udinger, Th. Seitz, H.-R. Trebin, and S. \v Zumer for
helpful discussions. H.~S. thanks P. Poulin, T.~C. Lubensky, and
D. Weitz for a fruitful collaboration, which initiated the present work. 
J.~S. gratefully acknowledges a grant from the Sonderforschungsbereich
382 of the Deutsche \linebreak

\noindent
Forschungsgemeinschaft.

\appendix
\section*{Method of Finite Elements for Nematic Liquid
Crystals in a Cylindrical Geometry}
The major advantage of the finite element method \cite{itapdb:Twizell1984}
is its ability to cope with arbitrarily complex geometries. In two 
dimensions, {\em e.g.}, the integration area is subdivided into 
{\em finite elements}, which in the simplest case are triangles.
%can be triangles or quadrilaterals.
In doing so, the boundaries of a complex geometry are well 
approximated by polygons. This is not possible within the method of
finite differences, where the grid is defined by the coordinate lines.
In our geometries strong director deformations occur close to the
boundaries of the spheres. For a quantitative analysis it was,
therefore, important to have these boundaries defined as smoothly as 
possible.

To determine the tilt angle field the 
Oseen-Z\"ocher-Frank free energy has to be minimized. 
The finite difference technique would directly address the 
Euler-Lagrange equation.
On the other hand, the strategy of the finite element method is to
start from a discretized version of the free energy
and then to apply a numerical minimization scheme. In our case,
the reduced free energy is written as a function
$\overline{F}[\bbox{\Theta}]$, where 
$\bbox{\Theta} = (\Theta_{1}, \ldots,\Theta_{n})$ denotes
the set of tilt angles on the $n$ vertices of the net.
It can be expressed as a sum over the free energies
for the individual finite elements:
\begin{equation}
\overline{F}[\bbox{\Theta}] = \sum_{j=1}^{l} 
\overline{F}_{j}[\Theta_{1}^{(j)},\Theta_{2}^{(j)},\Theta_{3}^{(j)}] \enspace,
\enspace \Theta_{1}^{(j)},\Theta_{2}^{(j)},\Theta_{3}^{(j)} \in \bbox{\Theta}
\enspace.
\end{equation}
The sum runs over all $l$ triangles, and 
$\overline{F}_{j}[\Theta_{1}^{(j)},\Theta_{2}^{(j)},\Theta_{3}^{(j)}]$ is
the free energy of the $j$-th triangle as a function of the
three tilt angles at the vertices of the triangle.
In order to derive 
$\overline{F}_{j}[\Theta_{1}^{(j)},\Theta_{2}^{(j)},\Theta_{3}^{(j)}]$
we start from the free energy of one triangle of area $A(j)$:
\begin{equation}
\overline{F}_{j}(\Theta) = 2\pi\,\int_{A(j)} \,\mbox{d}z\,\mbox{d}\rho\,
\rho\,\overline{f}[\Theta(\rho,z)] \enspace,
\end{equation}
The quantity $\overline{f}[\Theta(\rho,z)]$ is the 
Oseen-Z\"ocher-Frank free energy
density of Eq.\ (\ref{2}), and an axial symmetry is assumed.
Furthermore, in order to handle every finite element in the same manner,
we introduce {\em natural\/} coordinates $u,\,v$:
\begin{equation}
\begin{array}{l}
\rho^{(j)}(u,v) = \rho^{(j)}_1 
+ (\rho^{(j)}_2 - \rho^{(j)}_1)\,u + (\rho^{(j)}_3 - \rho^{(j)}_1)\,v \\
z^{(j)}(u,v) = z^{(j)}_1 + (z^{(j)}_2 - z^{(j)}_1)\,u 
+ (z^{(j)}_3 - z^{(j)}_1)\,v  \enspace,
\end{array}
\end{equation}
where $\rho^{(j)}_{i}$ and $z^{(j)}_{i}$ 
are the cylindrical coordinates of the
vertices of the $j$-th triangle ($i = 1, 2, 3$). 
In the new coordinates every triangle is right-angled and isosceles.
The free energy now assumes the form
\begin{eqnarray}
\overline{F}_{j}(\Theta) &=& 2\pi\,\Delta_{j} \,\times
\nonumber \\
& & \int_0^1\,\mbox{d}v\,\int_0^{1-v}\,\mbox{d}u\,\rho^{(j)}(u,v)\,
\overline{f}[\Theta^{(j)}(u,v)] \enspace,
\label{A3}
\end{eqnarray}
where
\begin{eqnarray}
\Delta_{j} &=& (\rho^{(j)}_2 - \rho^{(j)}_1)\,(z^{(j)}_3 - z^{(j)}_1) 
\nonumber \\
& & - (\rho^{(j)}_3 - \rho^{(j)}_1)\,(z^{(j)}_2 - z^{(j)}_1)
\label{A4}
\end{eqnarray}
is the Jacobian determinant of the coordinate transformation.
When the finite elements are small enough, the tilt angle field
within one element is well approximated by a linear interpolation
on the area of the triangle,
\begin{equation}
\label{A5}
\Theta^{(j)}(u,v) = \Theta^{(j)}_1 
+ (\Theta^{(j)}_2 - \Theta^{(j)}_1)\,u
+ (\Theta^{(j)}_3 - \Theta^{(j)}_1)\,v \enspace.
\end{equation}
We now insert (\ref{A5}) and (\ref{A4}) into (\ref{A3}). 
To simplify the integration over $u$ and $v$ we replace
$\Theta$ and $\rho$ by their average values 
$\rho^{(j)}_{0}=(\rho^{(j)}_{1}+\rho^{(j)}_{2}+\rho^{(j)}_{3})/3$ and 
$\Theta^{(j)}_{0} 
= (\Theta^{(j)}_{1} + \Theta^{(j)}_{2} + \Theta^{(j)}_{3}) / 3$.
This renders the integrals trivial, yielding the free energy of the
$j$-th triangle in terms of the three tilt angles at its three vertices:
\begin{equation}
\overline{F}_{j}[\Theta_{1}^{(j)},\Theta_{2}^{(j)},\Theta_{3}^{(j)}]
=  \pi\,\Delta_{j}\, \rho_{0}^{(j)}\,\overline{f}[\Theta^{(j)}] \enspace.
\end{equation}
The partial derivatives $\Theta_{\rho}$ and $\Theta_{z}$ 
necessary to compute $\overline{f}[\Theta^{(j)}]$ from
Eq.~(\ref{2}) follow from
\begin{eqnarray}
\Theta^{(j)}_{\rho} &=& \Theta^{(j)}_u\,u_{\rho}^{(j)} 
+ \Theta^{(j)}_v\,v_{\rho}^{(j)}, 
\\
\Theta^{(j)}_{z} &=& \Theta^{(j)}_u\,u_{z}^{(j)} 
+ \Theta^{(j)}_v\,v_{z}^{(j)}
\end{eqnarray}
with
\begin{equation}
\Theta^{(j)}_u = \Theta^{(j)}_2 - \Theta^{(j)}_1,\qquad
\Theta^{(j)}_v = \Theta^{(j)}_3 - \Theta^{(j)}_1,
\end{equation}
and
\begin{equation}
\begin{array}{rcl}
u_{\rho}^{(j)} &=& (z^{(j)}_3 - z^{(j)}_1)/\Delta_{j}, \\[.3ex]
v_{\rho}^{(j)} &=& - (z^{(j)}_2 - z^{(j)}_1)/\Delta_{j}, \\[.3ex]
u_{z}^{(j)} &=& - (\rho^{(j)}_3 - \rho^{(j)}_1) / \Delta_{j}, \\[.3ex]
v_{z}^{(j)} &=& (\rho^{(j)}_2 - \rho^{(j)}_1)/\Delta_{j} \enspace.
\end{array}
\end{equation}

%\bibliographystyle{prsty}
%\bibliographystyle{aip}
%\bibliography{jourkurz,/user6/holger/paper/emulnum1/paper}

\begin{thebibliography}{10}

\bibitem{itapdb:Kleman1983}
M. {Kl{\'e}man}, {\em Points, Lines and Walls: In liquid crystals, magnetic
  systems, and various ordered media} (John Wiley \& Sons, New York, 1983).

\bibitem{itapdb:Mermin1979}
N.~D. {Mermin}, Rev.\ Mod.\ Phys. {\bf 51},  591  (1979).

\bibitem{itapdb:Trebin1982}
H.-R. {Trebin}, Adv.\ Phys. {\bf 31},  195  (1982).

\bibitem{itapdb:Kurik1988}
M.~V. {Kurik} and O.~D. {Lavrentovich}, 
Usp. Fiz. Nauk \textbf{154}, 381 (1988),
[Sov.\ Phys.\ Usp. {\bf 31}, 196 (1988)].

\bibitem{itapdb:Chaikin1995}
P. {Chaikin} and T.~C. {Lubensky}, {\em Principles of Condensed Matter Physics}
  (Cambridge University Press, Cambridge, 1995).

\bibitem{itapdb:Schadt1971}
M. {Schadt} and W. {Helfrich}, Appl.\ Phys.\ Lett. {\bf 18},  127  (1971).

\bibitem{itapdb:Clark1980}
N. {Clark} and S. {Lagerwall}, Appl.\ Phys.\ Lett. {\bf 36},  899  (1980).

\bibitem{itapdb:Doane1986}
J.~W. {Doane}, N.~A. {Vaz}, B.~G. {Wu}, and S. {\v Zumer}, Appl.\ Phys.\ Lett.
  {\bf 48},  269  (1986).

\bibitem{itapdb:Drzaic1995}
P.~S. {Drzaic}, {\em Liquid {C}rystal {D}ispersions} (World Scientific Publ.,
  Singapore, 1995).

\bibitem{itapdb:Crawford1996}
{\em Liquid Crystals in Complex Geometries}, edited by G.~P. {Crawford} and S.
  {\v Zumer} (Taylor \& Francis, London, 1996).

\bibitem{itapdb:Russel1995}
W.~B. {Russel}, D.~A. {Saville}, and W.~R. {Schowalter}, {\em Colloidal
  Dispersions} (Cambridge University Press, Cambridge, 1995).

\bibitem{itapdb:Brochard1970}
F. {Brochard} and P.~G. de~{Gennes}, J.~Phys. {\bf 31},  691  (1970).

\bibitem{itapdb:Eidenschink1991}
R. {Eidenschink} and W.~H. de~{Jeu}, Electron.\ Lett. {\bf 27},  1195  (1991).

\bibitem{itapdb:Kreuzer1992}
M. {Kreuzer}, T. {Tschudi}, and R. {Eidenschink}, Mol.\ Cryst.\ Liq.\ Cryst.
  {\bf 223},  219  (1992).

\bibitem{itapdb:Glushchenko1997}
A. {Glushchenko} {\it et~al.}, Liq.\ Cryst. {\bf 23},  241  (1997).

\bibitem{itapdb:Poulin1994}
P. {Poulin}, V.~A. {Raghunathan}, P. {Richetti}, and D. {Roux}, J.~Phys.\ I\
  France {\bf 4},  1557  (1994).

\bibitem{itapdb:Raghunathan1996}
V.~A. {Raghunathan}, P. {Richetti}, and D. {Roux}, Langmuir {\bf 12},  3789
  (1996).

\bibitem{itapdb:Raghunathan1996a}
V.~A. {Raghunathan} {\it et~al.}, Mol.\ Cryst.\ Liq.\ Cryst. {\bf 288},  181
  (1996).

\bibitem{itapdb:Terentjev1995}
E.~M. {Terentjev}, Phys.\ Rev.~E {\bf 51},  1330  (1995).

\bibitem{itapdb:Kuksenok1996}
O.~V. {Kuksenok}, R.~W. {Ruhwandl}, S.~V. {Shiyanovskii}, and E.~M.
  {Terentjev}, Phys.\ Rev.~E {\bf 54},  5198  (1996).

\bibitem{itapdb:Ramaswamy1996}
S. {Ramaswamy}, R. {Nityananda}, V.~A. {Raghunathan}, and J. {Prost}, Mol.\
  Cryst.\ Liq.\ Cryst. {\bf 288},  175  (1996).

\bibitem{itapdb:Ruhwandl1997}
R.~W. {Ruhwandl} and E.~M. {Terentjev}, Phys.\ Rev.~E {\bf 55},  2958  (1997).

\bibitem{itapdb:Borstnik1997}
A. {Bor\v stnik} and S. {\v Zumer}, Phys.\ Rev.~E {\bf 56},  3021  (1997).

\bibitem{itapdb:Poulin1997}
P. {Poulin}, H. {Stark}, T.~C. {Lubensky}, and D.~A. {Weitz}, Science {\bf
  275},  1770  (1997).

\bibitem{itapdb:Poulin1998}
P. {Poulin} and D.~A. {Weitz}, Phys.\ Rev.~E {\bf 57},  626  (1998).

\bibitem{itapdb:Meyer1972}
R.~B. {Meyer}, Mol.\ Cryst.\ Liq.\ Cryst. {\bf 16},  355  (1972).

\bibitem{itapdb:Lubensky1998}
T.~C. {Lubensky}, D. {Pettey}, N. {Currier}, and H. {Stark}, Phys.\ Rev.~E {\bf
  57},  610  (1998).

\bibitem{itapdb:Poulin1997a}
P. {Poulin}, V. {Cabuil}, and D.~A. {Weitz}, Phys.\ Rev.\ Lett. {\bf 79},  4862
   (1997).

\bibitem{itapdb:Mermin1977}
N. {Mermin},  in {\em Quantum Fluids and Solids}, edited by S. {Trickey}, E.
  {Adams}, and J. {Dufty} (Plenum Press, New York, 1977).

\bibitem{itapdb:Candau1973}
S. {Candau}, P.~L. {Roy}, and F. {Debeauvais}, Mol.\ Cryst.\ Liq.\ Cryst. {\bf
  23},  283  (1973).

\bibitem{itapdb:Kurik1982}
M. {Kurik} and O. {Lavrentovich}, 
Pis'ma Zh. Eksp. Teor. Fiz. \textbf{35}, 65 (1982), 
[JETP Lett. {\bf 35},  65  (1982)].

\bibitem{itapdb:Twizell1984}
E.~H. {Twizell}, {\em Computational methods for partial differential equations}
  (Chichester, Horwood, 1984).

\bibitem{itapdb:deGennes1993}
P.~G. de~{Gennes} and J. {Prost}, {\em The Physics of Liquid Crystals}, {\em
  2nd. ed.} (Clarendon Press, Oxford, 1993).

\bibitem{itapdb:Press1992}
W.~H. {Press}, S.~A. {Teukolsky}, W.~T. {Vetterling}, and B.~P. {Flannery},
  {\em {N}umerical {R}ecipes in {F}ortran: {T}he {A}rt of {S}cientific
  {C}omputing} (Cambridge University Press, Cambridge, 1992).

\bibitem{itapdb:Landau1984f}
L.~D. {Landau} and E.~M. {Lifschitz}, {\em {S}tatistische {P}hysik, {T}eil 1},
  Vol.~5 of {\em {L}ehrbuch der {T}heoretischen {P}hysik}, 6th ed.
  (Akademie-Verlag, Berlin, 1984).

\end{thebibliography}

\end{document}